\documentclass[aps,twocolumn,showpacs,reprint,pra,amsmath,amssymb]{revtex4-1}

\usepackage{graphicx}
\usepackage{dcolumn}
\usepackage{bm}
\usepackage{amsmath}
\usepackage{color}

\allowdisplaybreaks[2]

\begin{document}

\title{Diffractionless image propagation and frequency conversion via four-wave mixing exploiting the  thermal motion of atoms}

\author{Lida Zhang}
\author{J\"{o}rg Evers}
\affiliation{Max-Planck-Institut f\"{u}r Kernphysik, Saupfercheckweg 1, D-69117 Heidelberg, Germany}

\date{\today}

\begin{abstract}
A setup to frequency-convert an arbitrary image encoded in the spatial profile of a probe field onto a signal field using four-wave mixing in a thermal atom vapor is proposed. The atomic motion is exploited to cancel diffraction of both signal and probe fields simultaneously. We show that an incoherent probe field can be used to enhance the transverse momentum bandwidth which can be propagated without diffraction, such that smaller structures with higher spatial resolution can be transmitted. It furthermore compensate linear absorption with non-linear gain, to improve the four-wave mixing performance since the propagation dynamics of the various field intensities is favorably modified.

\end{abstract}



\maketitle

\section{Introduction}      
A promising route towards the implementation of all-optical information processing or future quantum technologies is to combine different technologies into a hybrid system.  This way, the different core functionalities can be realized using different methods, exploiting their individual advantages.
But the various technologies typically operate at different characteristic frequencies, such that  efficient frequency conversion is usually required.
Such frequency conversion can be accomplished using non-linear four-wave mixing processes~\cite{shen}, which have already been widely studied and applied in various settings. Example are conventional up- and down-frequency conversion to generate vacuum-ultraviolet (VUV) or extreme-ultraviolet (XUV)~\cite{tewari1,harris1,tewari2,deng1,wu1} or far infrared (IR)~\cite{sorokin1,zibrov1,conversion,ding,liu} light.

But since FWM is a non-linear process, it sensitively depends on the intensities of the applied fields, and thereby on the propagation dynamics~\cite{4wm-dynamics} and on the transverse intensity profile of the applied fields. It is well-known, however, that light beams propagating in free space rapidly spread, together with a distortion of the transverse beam profile due to diffraction. The origin for diffraction is that each momentum component of the propagating beam acquires a different phase shift throughout the propagation. Diffraction is a fundamental limitation in particular for the creation, detection or propagation of small images, due to the large momentum bandwidth. The effect of diffraction thus is not only to destroy the information carried in the transverse beam profile, but also to distort any non-linear processes such as FWM, due to the modification of the spatial intensity profiles. This raises the question, whether diffraction can be manipulated such that non-linear processes can be effective. 

In order to suppress or even remove the diffraction, different methods have been suggested. First, there are particular characteristic spatial modes which satisfy the paraxial propagation equation, and therefore can propagate in free space without change in their transverse profile~\cite{durnin}. These are Airy~\cite{berry1979,siviloglou2007}, Bessel~\cite{durnin1987,nature2002,mcgloin2005}, Mathieu~\cite{vega2000,zhang2012} and parabolic (Weber) beams~\cite{zhang2012,bandres2004}. 
An alternative method is to induce a spatially-varying index of refraction experienced by the propagating light, effectively forming an optically written waveguide~\cite{truscott1999,agarwal2000,howell2009,dey2009,moseley,bortman,vengalatorre,tarhan,dey2011,zhang2013}. However, those schemes operating in position space typically only allow to propagate specific modes through a particular waveguide, and cannot be applied to multimode fields.

As an alternative approach, recently, Firstenberg et al. have proposed~\cite{firstenberg1,firstenberg2,firstenberg3} and demonstrated experimentally~\cite{firstenberg4} a novel scheme to eliminate paraxial diffraction of a probe field carrying an arbitrary image encoded into its spatial profile~\cite{rmp}. In contrast to previous approaches, it operates in the momentum space. The method takes advantage of electromagnetically induced transparency (EIT) in a thermal atomic vapor, under the influence of atomic motion and collisions. These induce a linear susceptibility experienced by the propagating field which is $\textbf{k}_{\bot}$-square dependent in momentum space. 
This susceptibility can be manipulated in such a way, that the phase shifts giving rise to diffraction are exactly canceled. Since the phase shift of each momentum component is eliminated individually, arbitrary images within a certain bandwidth can be propagated without diffraction. 
The method, however, necessarily involves strong single-photon absorption, since the diffraction can only be canceled at a  negative two-photon detuning between the probe and control fields, leading to an imperfect EIT condition.

\begin{figure}[t]
\centering
\includegraphics[width=6cm]{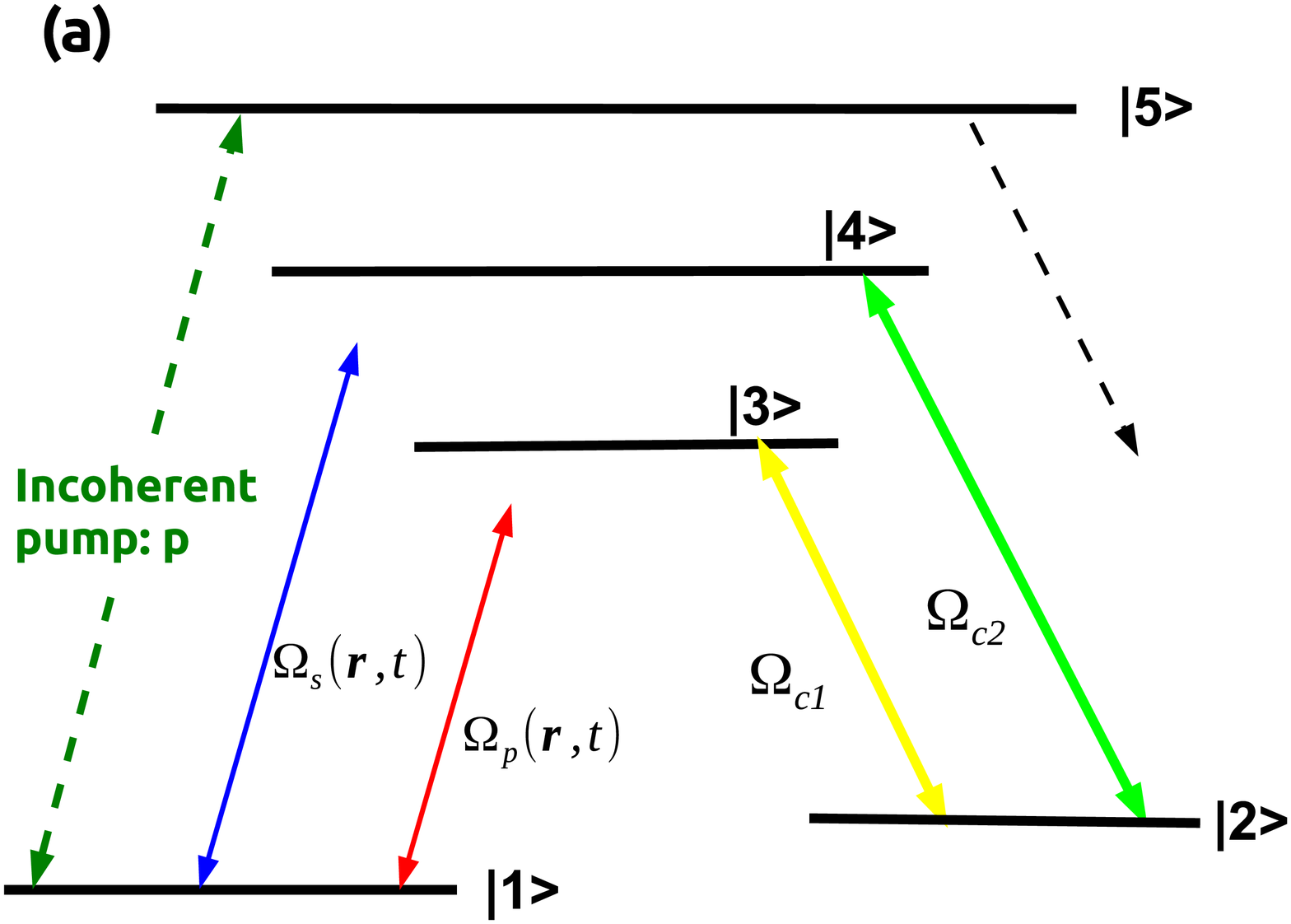}\\
\includegraphics[width=7cm]{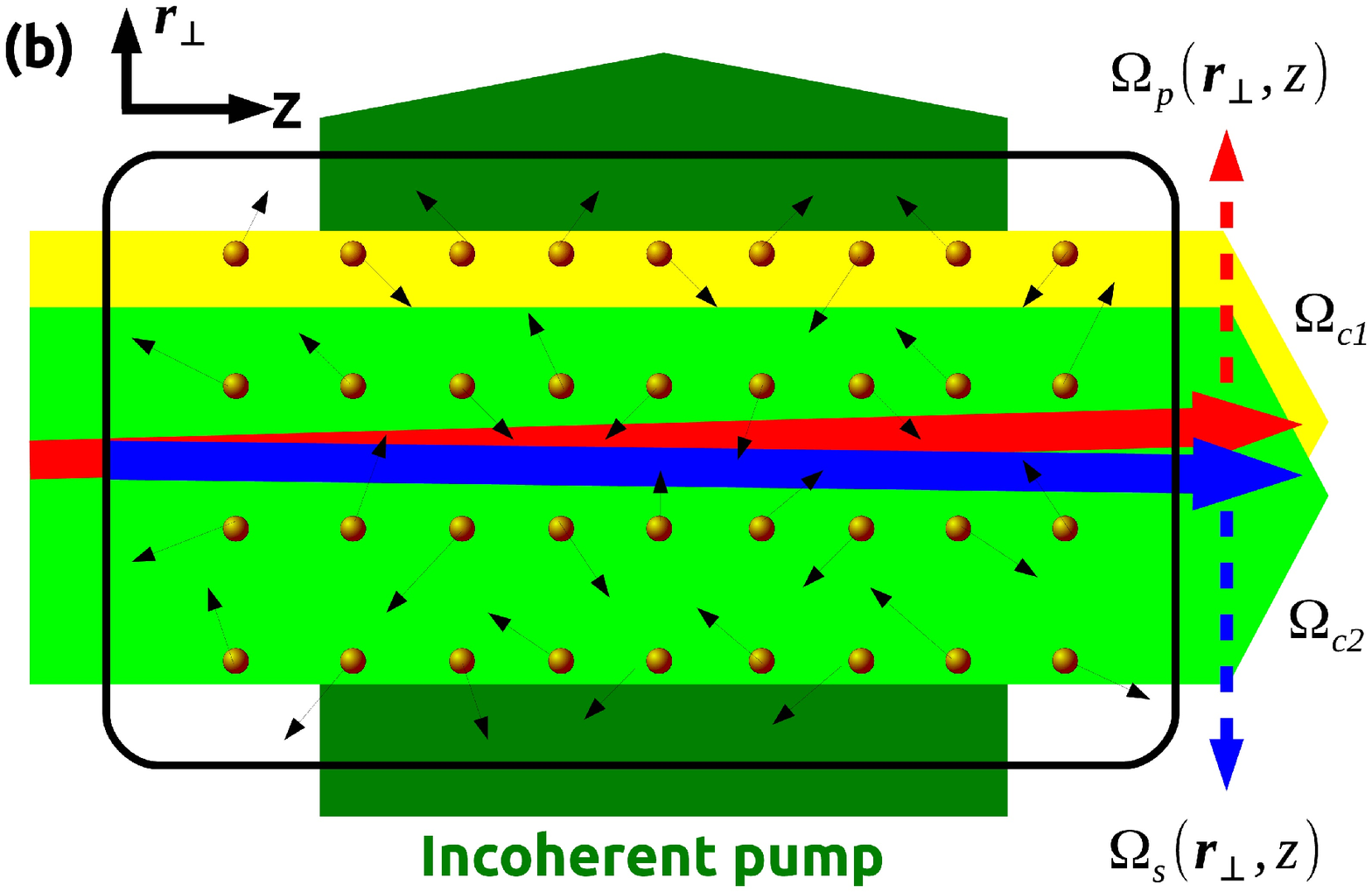}
\caption{(Color online) Schematic setup proposed to realize diffraction-less image reproduction via four-wave mixing in a thermal atom vapor. (a) In the considered five-level model, the lower three states in $\Lambda$-configuration form the basic electromagnetically-induced transparency setup for the probe propagation. The fourth state together with an additional control beam lead to a four-level double-$\Lambda$ system. which enables four-wave mixing to generate an additional signal field. Finally, a two-way incoherent pump field is applied to the fifth auxiliary state. In effect, the single-photon absorption for both the probe and signal fields are reduced, and the four-wave mixing process is enhanced. (b) Co-linear propagation of the four coherent fields in the thermal atomic medium to satisfy the phase matching condition. Due to atomic motion, the paraxial diffraction of arbitrary profiles of both probe and signal fields can be eliminated over a certain bandwidth of the transverse beam profile. 
Eventually, the incident image encoded onto the probe is copied to the FWM signal field, which also propagates without diffraction.
}
\label{fig:setup}
\end{figure}

Here, we propose a method to transfer and frequency-convert an arbitrary image encoded in the spatial profile of a probe field onto a signal field using four-wave mixing in a hot atom vapor. We demonstrate how diffraction can be avoided in this conversion, by exploiting the atomic motion to cancel the paraxial diffraction of both the probe and signal fields simultaneously. Furthermore, we show that an incoherent two-way pump field can be employed to improve the performance in various ways. It redistributes populations in zeroth order of the probe field, which together with the applied control fields leads to the formation of additional coherences. This on the one hand allows us to cancel linear absorption with  non-linear gain such that the light intensities are maintained throughout the propagation. On the other hand, the pump field broadens the transverse momentum bandwidth which can be propagated without diffraction, such that smaller spatial structures become accessible. Third, the pump leads to modified propagation dynamics of the field intensities which improves the FWM process, such that residual broadening and absorption of the probe and signal fields before the FWM equilibrium is reached are minimized. 
We analyze our setup first for Gaussian probe beams, and finally show that arbitrary images can be propagated and frequency-converted without diffraction. 
All examples are based on ${}^{87}$Rb at room temperature.

The paper is organized as follows. In Sec.~\ref{sec:setup}, we introduce our model and discuss a  possible experimental realization in ${}^{87}$Rb. In Sec.~\ref{sec:prop}, we characterize the spatial evolution of the probe and signal fields, and discuss the possibility to reproduce the spatial profile of the probe field onto the signal field. In Sec.~\ref{sec:susceptibilities}, we obtain the linear and nonlinear susceptibilities of the thermal atom vapor. 
Section~\ref{sec:result} discusses our main results. In Sec.~\ref{sec:sus}, we analyze the linear and non-linear susceptibilities, and discuss the elimination of diffraction and frequency conversion by the linear and nonlinear response of the medium. Results with and without incoherent pump fields are compared, and the effect of the pump is studied.
Sec.~\ref{sec:dynamics} presents our numerical results on the propagation dynamics of a pump field with Gaussian spatial profile. We show that it can be frequency converted into a signal field via the FWM process, and that the two resulting fields can be propagated without diffraction. The effects of the incoherent pump on the power and width of the outgoing probe and signal fields are studied in detail in Sec.~\ref{sec:pumpdynamics}. Finally, we show in Sec.~\ref{sec:2d} that a two-dimensional image encoded in the spatial profile of the  probe field can be propagated without diffraction and is reproduced to the FWM signal field in a suitably prepared thermal medium. Sec.~\ref{sec:summary} concludes with a discussion and summary. Details on the calculation of the linear and nonlinear susceptibilities including the effect of the thermal motion are provided in the Appendix~\ref{sec:appendix}.

\section{Theoretical Model}
\subsection{\label{sec:setup}Setup}

We consider a five-level atomic system as shown in Fig.~\ref{fig:setup}(a), which is designed as follows. The three levels $|1\rangle$, $|2\rangle$ and $|3\rangle$ form a $\Lambda$-shaped electromagnetically induced transparency setup. Transition $|1\rangle\leftrightarrow|3\rangle$ is driven by a weak probe field with Rabi frequency $\Omega_p(\textbf{r},t)=\vec\mu_{31}\cdot\vec e_{p}E_{p}(\textbf{r},t)/2\hbar$, while a resonant control field with Rabi frequency $\Omega_{c1}(\textbf{r},t)=\vec\mu_{32}\cdot\vec e_{c1}E_{c1}(\textbf{r},t)/2\hbar$ is applied to transition $|2\rangle\leftrightarrow|3\rangle$.
This $\Lambda$-type setup is enlarged to a double-$\Lambda$-system by additionally applying a resonant control field with Rabi frequency  $\Omega_{c2}(\textbf{r},t)=\vec\mu_{42}\cdot\vec e_{c2}E_{c2}(\textbf{r},t)/2\hbar$ to transition $|2\rangle\leftrightarrow|4\rangle$. 
The fourth transition $|1\rangle\leftrightarrow|4\rangle$ is not driven by an external field, but due to four-wave mixing, a signal field $\Omega_s(\textbf{r},t)=\vec\mu_{41}\cdot\vec e_{s}E_{s}(\textbf{r},t)/2\hbar$ can be generated throughout the propagation of the light fields through the medium.
Finally, we apply an incoherent two-way pump field to the transition $|1\rangle\leftrightarrow|5\rangle$~\cite{zibrov2,kapale1,wang1,obrien1}, which together with spontaneous decay from $|5\rangle$ into the other states effectively forms a one-way pumping from $|1\rangle$ into the other atomic states.
Here, $\vec\mu_{ij}$ is the dipole moment between states $|i\rangle$ and $|j\rangle$ ($i,j\in\{1,2,3,4\}$).  $E_{n}$ and $\vec e_{n}$ are the slowly varying envelopes and unit polarization vectors of the electric fields with $n\in\{p,s,c1,c2\}$.
The incoherent pump field induces a redistribution of the populations in zeroth order of the probe field, which together with the control fields leads to the formation of additional atomic coherences. These coherences will turn out to favorably affect the propagation dynamics in a non-trivial way.
 
As discussed in more detail later, the required level scheme can be realized, e.g., in $^{87}\text{Rb}$, with magnetic sublevels 5$^{2}S_{1/2}, F=1,m_{F}=-1$ and $F=2,m_{F}=-2$ as the two lower states $|1\rangle$ and  $|2\rangle$, and 5$^{2}P_{1/2}, F=2,m_{F}=-1$ and $F=1,m_{F}=-1$ as the two upper states $|3\rangle$ and $|5\rangle$, respectively. State $|4\rangle$ then can be chosen as 5 $^{2}P_{3/2}, F=2,m_{F}=-2$. We use the parameters of this implementation as summarized in the caption of Fig.~\ref{fig:lin} for the numerical calculations presented below.

\subsection{\label{sec:prop}Propagation dynamics}

In order to derive analytical expressions, we assume the undepleted pump approximation for the strong control fields, such that they only acquire phase shifts as result of self- and cross- phase modulation during the FWM process. In paraxial approximation for propagation in $z$ direction, we obtain the propagation equations in momentum space
\begin{subequations}
\label{eq3} 
\begin{align}
&\left(\frac{\partial}{\partial z}+i\frac{\text{k}_{\bot}^{2}}{2\text{k}_{p}}\right)\Omega_{p}(\textbf{k}_{\bot},z) \nonumber \\
&=i\frac{\text{k}_{p}}{2}\left [\chi_{p}(\textbf{k}_{\bot})\Omega_{p}(\textbf{k}_{\bot},z)+\chi_{sp}(\textbf{k}_{\bot})\Omega_{s}(\textbf{k}_{\bot},z)\right] \,,\\
&\left(\frac{\partial}{\partial z}+i\frac{\text{k}_{\bot}^{2}}{2\text{k}_{s}}\right)\Omega_{s}(\textbf{k}_{\bot},z) \nonumber \\
&=i\frac{\text{k}_{s}}{2}\left [\chi_{s}(\textbf{k}_{\bot})\Omega_{s}(\textbf{k}_{\bot},z)+\chi_{ps}(\textbf{k}_{\bot})\Omega_{p}(\textbf{k}_{\bot},z)\right ]\,.
\end{align}
\end{subequations}
Here, $\text{k}_{p}$ [$\text{k}_{s}$] is the wave number of the probe [signal] field. The terms $ik_{\bot}^{2}/2k_{p(s)}$ characterize the paraxial diffraction of the probe (signal) field, and lead to spatial broadening and energy spreading throughout the field propagation, thereby severely distorting the spatial profiles of the incident fields. The linear response of the atomic medium to the probe and signal fields are given by $\chi_{p}(\textbf{k}_{\bot})$ and $\chi_{s}(\textbf{k}_{\bot})$ respectively, while $\chi_{ps}(\textbf{k}_{\bot})$ [$\chi_{sp}(\textbf{k}_{\bot})$] characterizes the nonlinear forward [backward] FWM process from $\Omega_p$ [$\Omega_s$] to $\Omega_s$ [$\Omega_p$]. 

We can see from Eqs.~(\ref{eq3}) that each wave vector component of the probe field is proportionally transfered onto the signal field and vice versa. Therefore, eventually images carried by the transverse degrees of freedom of the probe field can be copied to the signal field via the interplay of the FWM processes. In this transfer, the relative intensities of the output probe and signal fields depend on their coupling strengths to the atomic medium. As shown in the following, the combination of the linear and nonlinear responses of the thermal vapor can be manipulated in such a way that it exactly eliminates the diffraction of both fields due to atomic motion when the FWM process reaches the equilibrium. As a result, the initial spatial profile of the probe field is then reproduced and thereby frequency converted  essentially without diffraction to the signal field during the FWM process.

\subsection{\label{sec:susceptibilities}Linear and non-linear susceptibilities}

For the calculation of the linear susceptibilities $\chi_p,\chi_s$ and the nonlinear susceptibilities $\chi_{ps},\chi_{sp}$ under the influence of atomic motion, we follow the approach introduced in Ref.~\cite{firstenberg2}. In the following, we focus on the main results, whereas the detailed derivation is summarized in the Appendix. Throughout the derivation, we apply a sequence of approximations.  In particular, we assume that the two control fields and the incoherent pump are plane waves. For the weak probe and signal fields, we apply the slowly-varying envelope and paraxial approximations, but allow for arbitrary spatial profiles within the paraxial regime. In the Dicke limit, in which the Doppler shift for the Raman two-photon transition $\Delta\text{k}\text{v}_{\text{th}}$ is much smaller than the combination of the collision rate $\gamma_{c}$ and the incoherent pump $p$, i.e.,
\begin{align}
\label{dickelimit}
\Delta\text{k}\text{v}_{\text{th}}\ll\gamma_{c}+\frac{p}{2}\,,
\end{align}
we find  
\begin{subequations}
\label{susceptibilities} 
\begin{align}
\chi_{p}(\textbf{k}_{\bot})&=i\frac{3\lambda^{3}_{p}K_{31}n_{0}\Gamma_{31}}{8\pi^{2}} 
\left(\rho^{(0)}_{11}-\rho^{(0)}_{33} \right. \nonumber \\
&\left.+\frac{\Gamma_{c1}(\rho^{(0)}_{11}-\rho^{(0)}_{33})+i\Omega_{c1}\rho^{(0)}_{23}-\Gamma_{b}\rho^{(0)}_{43}}{i\Delta-\Gamma_{1}-D\text{k}_{\bot}^2}\right),\\[2mm]
\chi_{s}(\textbf{k}_{\bot})&=i\frac{3\lambda^{3}_{s}K_{41}n_{0}\Gamma_{41}}{8\pi^{2}} 
\left(\rho^{(0)}_{11}-\rho^{(0)}_{44}\right. \nonumber \\ 
&\left.+\frac{\Gamma_{c2}(\rho^{(0)}_{11}-\rho^{(0)}_{44})+i\Omega_{c2}\rho^{(0)}_{24}-\Gamma_{a}\rho^{(0)}_{34}}{i\Delta-\Gamma_{1}-D\text{k}_{\bot}^2}\right),\\[2mm]
\chi_{sp}(\textbf{k}_{\bot})&=i\frac{3\lambda^{3}_{p}K_{31}n_{0}\Gamma_{31}}{8\pi^{2}} 
\left(-\rho^{(0)}_{34}\right. \nonumber \\
&\left.+\frac{\Gamma_{b}(\rho^{(0)}_{11}-\rho^{(0)}_{44})+i\Omega_{c1}\rho^{(0)}_{24}-\Gamma_{c1}\rho^{(0)}_{34}}{i\Delta-\Gamma_{1}-D\text{k}_{\bot}^2}\right),\\[2mm]   
\chi_{ps}(\textbf{k}_{\bot})&=i\frac{3\lambda^{3}_{s}K_{41}n_{0}\Gamma_{41}}{8\pi^{2}} 
\left(-\rho^{(0)}_{43} \right. \nonumber \\
&\left.+\frac{\Gamma_{a}(\rho^{(0)}_{11}-\rho^{(0)}_{33})+i\Omega_{c2}\rho^{(0)}_{23}-\Gamma_{c2}\rho^{(0)}_{43}}{i\Delta-\Gamma_{1}-D\text{k}_{\bot}^2}\right),
\end{align}
\end{subequations}
where $\lambda_{p}$ [$\lambda_{s}$] is the wavelength of the probe [signal] field, and $n_{0}$ is the atomic density. $K_{31}$ and $K_{41}$ are related to the single-photon spectrum for both fields and can be treated as real numbers close to resonance~\cite{firstenberg2}. Their definitions are given in Eqs.~(\ref{eq17}) of the appendix. $\Delta=\Delta_{p}-\Delta_{c1}$ is the two-photon detuning of the lower $\Lambda$-subsystem, $\Gamma_{ij}$ is the spontaneous emission rate from state $|i\rangle$ to $|j\rangle$ and $\rho^{(0)}_{ij}$ are the zeroth order populations or coherences for a single atom at rest, which are determined by the incoherent pump $p$ and the two control fields $\Omega_{c1}$ and $\Omega_{c2}$. Further, 
\begin{align}
D&=\frac{\text{v}^{2}_{\text{th}}}{\gamma_{c1}-i\Delta}\,, \label{eqd}\\
\gamma_{c1}&=\gamma_{c}+p/2+\gamma_{21}\label{eqc1}\,.
\end{align}
The effective line width in the susceptibilities Eqs.~(\ref{susceptibilities}) is
\begin{align}
\label{lw}
 \Gamma_{1}&=p/2 +\gamma_{21}+\Gamma_{c1}+\Gamma_{c2}\,,
\end{align}
with power broadening contributions of the two control fields $\Gamma_{c1}=K_{31}\Omega^{2}_{c1}$ and $\Gamma_{c2}=K_{41}\Omega^{2}_{c2}$, and $\Gamma_{a}=K_{31}\Omega^{\ast}_{c1}\Omega_{c2}$ and $\Gamma_{b}=K_{41}\Omega_{c1}\Omega^{\ast}_{c2}$.

\section{\label{sec:result}Results}
\subsection{\label{sec:sus}Susceptibilities}

\subsubsection{\label{without}Without incoherent pump}

In this section, we show how the atomic motion can be exploited to eliminate the paraxial diffraction of both probe and signal fields. We start with the case without incoherent pump field, i.e., $p=0$. Then, $\rho^{(0)}_{11}=1$ is the only non-vanishing zeroth-order density matrix element, and Eqs.~(\ref{susceptibilities}) simplify to
\begin{subequations}
\label{eq20} 
\begin{align}
\chi_{p}(\textbf{k}_{\bot})&=i\frac{3\lambda^{3}_{p}K_{31}n_{0}\Gamma_{31}}{8\pi^{2}} 
\left(1+\frac{\Gamma_{c1}}{i\Delta-\Gamma_0-D_{0}\text{k}_{\bot}^2}\right)\,,\\
\chi_{s}(\textbf{k}_{\bot})&=i\frac{3\lambda^{3}_{s}K_{41}n_{0}\Gamma_{41}}{8\pi^{2}} 
\left(1+\frac{\Gamma_{c2}}{i\Delta-\Gamma_0-D_{0}\text{k}_{\bot}^2}\right)\,,\\
\chi_{sp}(\textbf{k}_{\bot})&=i\frac{3\lambda^{3}_{p}K_{41}n_{0}\Gamma_{31}}{8\pi^{2}} 
\frac{\Gamma_{b}}{i\Delta-\Gamma_0-D_{0}\text{k}_{\bot}^2} \,,\\ 
\chi_{ps}(\textbf{k}_{\bot})&=i\frac{3\lambda^{3}_{s}K_{41}n_{0}\Gamma_{41}}{8\pi^{2}} 
\frac{\Gamma_{a}}{i\Delta-\Gamma_0-D_{0}\text{k}_{\bot}^2}\,,
\end{align}
\end{subequations}
where $\Gamma_0=\gamma_{21}+\Gamma_{c1}+\Gamma_{c2}$ and $D_{0}=\text{v}^{2}_{\text{th}}/(\gamma_{c}+\gamma_{21}-i\Delta)$. Setting $\Omega_{c2}=0$ such that state $|4\rangle$ is not accessed, the system reduces to a 3-level $\Lambda$-type setup, and we recover the result for $\chi_{p}(\textbf{k}_{\bot})$ of Ref.~\cite{firstenberg2,firstenberg4}. 
In the Dicke limit, Eqs.~(\ref{eq20}) can be approximated to first order in $\text{k}^{2}_{\bot}$ to give
\begin{subequations}
\label{eq21} 
\begin{align}
&\chi_{p}(\textbf{k}_{\bot})=\frac{3\lambda^{3}_{p}K_{31}n_{0}\Gamma_{31}}{8\pi^{2}} 
\left[i\left(1-\frac{\Gamma_{c1}}{2\Gamma_0}\right)-\frac{\Gamma_{c1}}{2\Gamma_0}\left(1-\frac{\text{k}^{2}_{\bot}}{\text{k}^2_{0}}\right)\right]\,,\\
&\chi_{s}(\textbf{k}_{\bot})=\frac{3\lambda^{3}_{s}K_{41}n_{0}\Gamma_{41}}{8\pi^{2}} 
\left[i\left(1-\frac{\Gamma_{c2}}{2\Gamma_0}\right)-\frac{\Gamma_{c2}}{2\Gamma_0}\left(1-\frac{\text{k}^{2}_{\bot}}{\text{k}^2_{0}}\right)\right]\,,\\
&\chi_{sp}(\textbf{k}_{\bot})=\frac{3\lambda^{3}_{p}K_{41}n_{0}\Gamma_{31}}{8\pi^{2}} 
\frac{\Gamma_{b}}{2\Gamma_0}\left[-1-i+\frac{\text{k}^{2}_{\bot}}{\text{k}^2_{0}}\right]\,,\\
&\chi_{ps}(\textbf{k}_{\bot})=\frac{3\lambda^{3}_{s}K_{41}n_{0}\Gamma_{41}}{8\pi^{2}} 
\frac{\Gamma_{a}}{2\Gamma_0}\left[-1-i+\frac{\text{k}^{2}_{\bot}}{\text{k}^2_{0}}\right]\,.
\end{align}
\end{subequations}
Here, $\text{k}_{0}=\sqrt{\Gamma_0/D_{\text{r0}}}$ with $D_{\text{r0}}=\text{v}^2_{\text{th}}/\gamma_{c}$, where we have neglected $\gamma_{21}$ since it is much smaller than $\gamma_{c}$. We have also set $\Delta=-\Gamma_0$ in order to remove the dependence of the imaginary parts of the linear and nonlinear susceptibilities on $\text{k}^{2}_{\bot}$. Note that the real parts proportional to $\text{k}^{2}_{\bot}$ remain positive in the central region $\text{k}_{\bot}\ll\text{k}_{0}$. 

\begin{figure}[t]
\centering
\includegraphics[width=0.8\columnwidth]{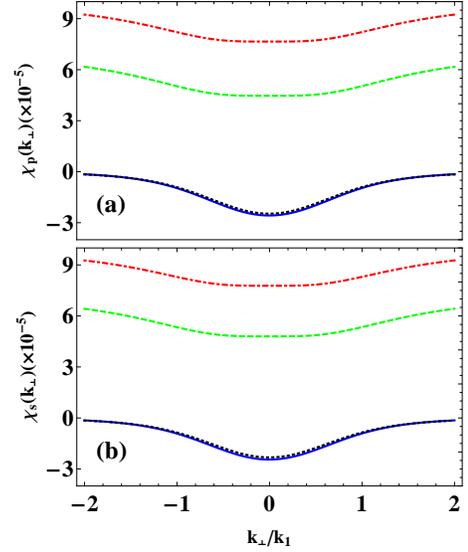}
\caption{(Color online) Motion-induced linear susceptibilities for the probe (a) and signal (b) fields in momentum space.  
The figures show the linear dispersion without incoherent pump field (black dotted line) and with incoherent pump field (solid blue curve). It can be seen that the dispersion is essentially unaffected by the pump and remains approximately quadratic in the central region $\textbf{k}_{\bot}\ll\textbf{k}_{1}$. In contrast, the linear absorption with (dashed green curve) and without (dot-dashed red curve) are strongly modified by the incoherent pump. 
Parameters are $\lambda_{p}=795\text{nm}$, $\lambda_{s}=780\text{nm}$, $\Gamma_{\text{D1}}=2\pi\times 5.75 \text{MHz}$, $\Gamma_{\text{D2}}=2\pi\times 6.07 \text{MHz}$, $\Gamma_{31}=\Gamma_{\text{D1}}/4$,$\Gamma_{32}=\Gamma_{\text{D1}}/6$, $\Gamma_{41}=\Gamma_{\text{D2}}/4$, $\Gamma_{42}=\Gamma_{\text{D2}}/6$, $\Gamma_{51}=\Gamma_{\text{D1}}/12$, $\Gamma_{52}=\Gamma_{\text{D2}}/2$, $\gamma_{21}=0.001\Gamma_{31}$, $T=300$K, $\text{v}_{\text{th}}=240 \text{m/s}$, $\Delta\text{k}=22.8\text{m}^{-1}$, $\Omega_{c1}=1.55\Gamma_{32}$, $\Omega_{c2}=1.43\Gamma_{42}$, $\Delta_{c1}=0$, $\Delta_{c2}=0$, and $\Delta_{p}=\Delta$.
With incoherent pump field, $n_{0}=1.32\times 10^{18}\text{m}^{-3}$, $\gamma_{c}=1600\Delta\text{k}\cdot\text{v}_{\text{th}}$, and $p=0.7\Gamma_{31}$. Without incoherent pump field, $n_{0}=6.2\times 10^{17}\text{m}^{-3}$, $\gamma_{c}=30000\Delta\text{k}\cdot\text{v}_{\text{th}}$, and $p=0$. In both cases,  parameters are chosen such that the diffraction for both probe and signal fields can be eliminated, and that the transverse momentum scales with and without pump are comparable,  $\text{k}_{0}\approx\text{k}_{1}$.} 
\label{fig:lin}
\end{figure}

\begin{figure}[t]
\centering
\includegraphics[width=0.8\columnwidth]{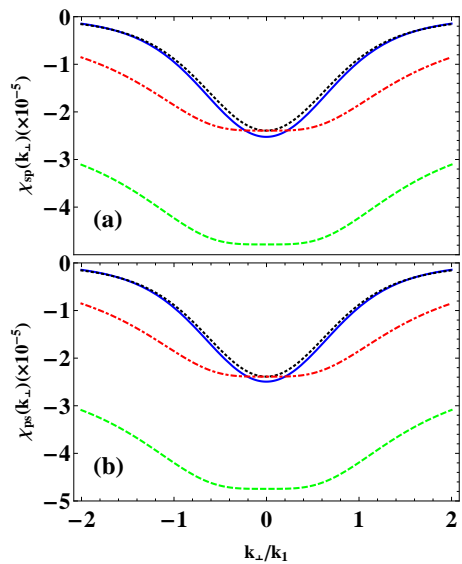}
\caption{(Color online) Motion-induced non-linear susceptibilities related to the forward (a) and backward (b) FWM processes. 
The figures show the non-linear dispersion without incoherent pump field (black dotted line) and with incoherent pump field (solid blue curve). As for the linear susceptibilities, the dispersion is essentially unaffected by the pump and remains approximately quadratic in the central region $\textbf{k}_{\bot}\ll\textbf{k}_{1}$. In contrast, the non-linear absorption with (dashed green curve) and without (dot-dashed red curve) is strongly modified by the incoherent pump. Parameters can be chosen such that the non-linear and the linear absorption cancel for a suitable choice of the pump field.
Parameters are as in Fig.~\ref{fig:lin}.}
\label{fig:nonlin}
\end{figure}

Results for the linear and nonlinear susceptibilities in Eq.~(\ref{eq20}) as a function of $\text{k}_{\bot}$ are shown in Figs.~\ref{fig:lin} and~\ref{fig:nonlin}. Note that in order to facilitate a comparison of the shapes of the respose curves, in these figures, we have chosen parameters in such a way that $\text{k}_{0}\approx\text{k}_{1}$, i.e., the scales of the transverse wave vectors with and without incoherent pump fields are approximately the same.
In the central area $\text{k}_{\bot}\ll\text{k}_{1}$, the linear and nonlinear dispersions are quadratic in $\text{k}_{\bot}$ with a constant offset (black dotted line). Recalling the propagation equations in paraxial approximation for the probe and signal fields Eqs.~(\ref{eq3}), we immediately find that the combination of linear and nonlinear susceptibility has a suitable functional dependence to cancel the effect of diffraction throughout the propagation, provided that the two fields have the same spatial structure. Indeed, as shown later, the FWM process duplicates the spatial shape of the probe onto the signal field. Since the elimination of diffraction operates over a certain momentum range, it does not depend on the spatial profile of both fields within a certain bandwidth range. Within this bandwidth, diffraction of arbitrary spatial 
structure can be removed by 
the atomic motion~\cite{firstenberg2}. 

This elimination of the diffraction, however, is accompanied by strong single-photon absorption as shown in Fig.~\ref{fig:lin} by the dot-dashed red curve. This significant attenuation of  the output intensities of both fields is inevitable without pump field, since a negative two-photon detuning $\Delta=-\Gamma_0$ deviating from the EIT resonance condition is required for the cancellation of diffraction. This forms a major obstacles for practical applications of the present scheme.

\begin{figure}[t]
\centering
\includegraphics[width=0.8\columnwidth]{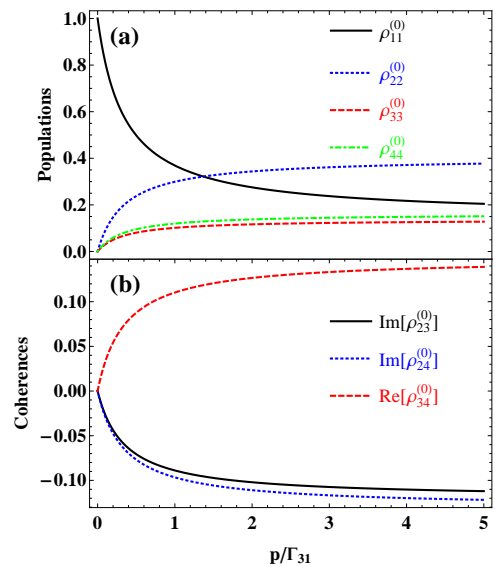}
\caption{\label{fig:pop}(Color online) (a) Zeroth-order populations $\rho^{(0)}_{ii}(i=1,2,3,4)$ and (b) coherences $\rho^{(0)}_{ij}(i\neq j)$ for a motionless atom as a function the intensity of the two-way incoherent pump field. At zero pump intensity, the atoms are in the ground state $|1\rangle$. As the intensity of the incoherent pump increases, atoms are gradually redistributed among the other four states, and atomic coherences arise between $|2\rangle$, $|3\rangle$ and $|4\rangle$ due to interactions with the two coherent control fields $\Omega_{c1}$ and $\Omega_{c2}$. Parameters are as in Fig.~\ref{fig:lin}. } 
\end{figure}

\subsubsection{With incoherent pump}

We next consider the effect of the incoherent pump field, which redistributes the populations in the five-level system. Subsequently, the control fields create additional coherences already at zeroth order in the probe and signal fields. The aim of this control is to cancel the single-photon absorption, without perturbing the diffraction cancellation. The zero-order populations and coherences $\rho^{(0)}_{ij}$ for motionless atoms can be obtained by solving the steady-state master equation. The analytic exact expressions for $\rho^{(0)}_{ij}$ are too complicated to be shown here. Instead, we show the relevant $\rho^{(0)}_{ij}$ as a function of the incoherent pump field strength in Fig.~\ref{fig:pop}. As the intensity of the incoherent pump increases, the population of the initial ground state $|1\rangle$ decreases, while that of the other states increases. Due to the two coherent control fields, atomic coherences are established in the control field part of the double-$\Lambda$ system. For resonant control 
fields $\Delta_{c1}=\Delta_{c2}=0$ as considered in the following, $\rho^{(0)}_{23}$ and $\rho^{(0)}_{24}$ are purely imaginary while $\rho^{(0)}_{34}$ is  real. 

Assuming again the Dicke limit, we can expand Eqs.~(\ref{susceptibilities}) to first order in $\text{k}^{2}_{\bot}$ to give $(i\in\{s,p,sp,ps\})$
\begin{align}
\chi_{i}(\textbf{k}_{\bot})\approx\chi^{(0)}_{i}+\chi^{(1)}_{i}\frac{\text{k}^{2}_{\bot}}{\text{k}^{2}_{1}}+O(\text{k}^{4}_{\bot})  \,.
\label{expand}
\end{align}
Specifically, we obtain
\begin{subequations}
\label{eq23} 
\begin{align}
\chi_{p}(\textbf{k}_{\bot})&=\frac{3\lambda^{3}_{p}K_{31}n_{0}\Gamma_{31}}{8\pi^{2}} 
\bigg\{i(\rho^{(0)}_{11}-\rho^{(0)}_{33}) \nonumber\\
&+\frac{\Gamma_{c1}(\rho^{(0)}_{11}-\rho^{(0)}_{33})+i\Omega_{c1}\rho^{(0)}_{23}-\Gamma_{b}\rho^{(0)}_{43}}{\Gamma_{1}}\nonumber\\
&\times \big[-\frac{1+i}{1+\alpha^{2}}+\frac{\alpha(2\Gamma_{1}+\gamma_{c1})^{2}}{2(\gamma_{c1}+\Gamma_{1})^2}\frac{\text{k}^{2}_{\bot}}{\text{k}^{2}_{1}}\big]\bigg\}\,,\\[1ex]
\chi_{s}(\textbf{k}_{\bot})&=\frac{3\lambda^{3}_{s}K_{31}n_{0}\Gamma_{41}}{8\pi^{2}} 
\bigg\{i(\rho^{(0)}_{11}-\rho^{(0)}_{44}) \nonumber\\
&+\frac{\Gamma_{c2}(\rho^{(0)}_{11}-\rho^{(0)}_{44})+i\Omega_{c2}\rho^{(0)}_{24}-\Gamma_{a}\rho^{(0)}_{34}}{\Gamma_{1}} \nonumber\\
&\times \big[-\frac{1+i}{1+\alpha^{2}}+\frac{\alpha(2\Gamma_{1}+\gamma_{c1})^{2}}{2(\gamma_{c1}+\Gamma_{1})^2}\frac{\text{k}^{2}_{\bot}}{\text{k}^{2}_{1}}\big]\bigg\} \,,\\[1ex]
\chi_{sp}(\textbf{k}_{\bot})&=\frac{3\lambda^{3}_{p}K_{41}n_{0}\Gamma_{31}}{8\pi^{2}} 
\bigg\{-i\rho^{(0)}_{34} \nonumber\\
&+\frac{\Gamma_{b}(\rho^{(0)}_{11}-\rho^{(0)}_{44})+i\Omega_{c1}\rho^{(0)}_{24}-\Gamma_{c1}\rho^{(0)}_{34}}{\Gamma_{1}} \nonumber\\
&\times \big[-\frac{1+i}{1+\alpha^{2}}+\frac{\alpha(2\Gamma_{1}+\gamma_{c1})^{2}}{2(\gamma_{c1}+\Gamma_{1})^2}\frac{\text{k}^{2}_{\bot}}{\text{k}^{2}_{1}}\big]\bigg\} \,, \\[1ex]  
\chi_{ps}(\textbf{k}_{\bot})&=\frac{3\lambda^{3}_{s}K_{41}n_{0}\Gamma_{41}}{8\pi^{2}} 
\bigg\{-i\rho^{(0)}_{43} \nonumber \\
&+\frac{\Gamma_{a}(\rho^{(0)}_{11}-\rho^{(0)}_{33})+i\Omega_{c2}\rho^{(0)}_{23}-\Gamma_{c2}\rho^{(0)}_{43}}{\Gamma_{1}} \nonumber\\
&\times \big[-\frac{1+i}{1+\alpha^{2}}+\frac{\alpha(2\Gamma_{1}+\gamma_{c1})^{2}}{2(\gamma_{c1}+\Gamma_{1})^2}\frac{\text{k}^{2}_{\bot}}{\text{k}^{2}_{1}}\big]\bigg\} \,.
\end{align}
\end{subequations}
Here,
\begin{align}
\alpha = \sqrt{\frac{\gamma_{c1}}{2\Gamma_{1}+\gamma_{c1}}}\,,
\end{align}
and the critical transverse wave number scale is set by
\begin{align}
\label{k1}
\text{k}_{1}=\sqrt{\frac{\Gamma_{1}}{D_{\text{r1}}}}=\text{v}_{\text{th}}\sqrt{\left( \frac{p}{2}+\Gamma_{c1}+\Gamma_{c2}\right) \left(\frac{p}{2}+\gamma_{c}\right )}\,,
\end{align}
where $D_{\text{r1}}=\text{v}^2_\text{th}/\gamma_{c1}$ follows from Eq.~(\ref{eqd}) on resonance and $\Gamma_{1}$ is defined in Eq.~(\ref{lw}). We find that $\text{k}_{1}$ grows rapidly with increasing  incoherent pump rate $p$, which enables operation of the diffraction-less image propagation over a larger transverse wave number bandwidth compared to the case without incoherent pump field.

In order to remove the dependence on $\text{k}^{2}_{\bot}$ for the imaginary parts of the linear and nonlinear susceptibilities and at the same time keep the real parts positive in the central area $\text{k}_{\bot}\ll\text{k}_{1}$ as desired, we have calculated the condition for the two-photon detuning
\begin{align}
\Delta =-\alpha\Gamma_{1}\,.
\label{eq24}
\end{align}

In Figs.~\ref{fig:lin} and~\ref{fig:nonlin}, we show the imaginary and real parts of the linear and nonlinear susceptibilities against the transverse wave vector $\text{k}_{\bot}$. It can be seen that the single-photon absorption has been considerably reduced due to the incoherent pump (dashed green curve), together with an increase of the non-linear gain. Still, both the linear and the non-linear absorption remain approximately constant in the central area $\text{k}_{\bot}\ll\text{k}_{1}$. At the same time, the linear and nonlinear dispersions (solid blue curve) are proportional to $\text{k}^{2}_{\bot}$ except for constant offsets, and essentially unchanged compared to the case without incoherent pumping. 
We thus conclude that the parameters can be chosen such that the paraxial diffraction of the both probe and the signal fields can be approximately eliminated after the FWM process has reached its equilibrium,  but with significant reduction of absorption compared to the case without pump. 

It should be noted that it is not possible to fully compensate the single-photon absorption for both signal and probe fields using only the linear gain from the atomic coherences, since $\rho^{(0)}_{43}$  [$\rho^{(0)}_{34}$] in $\chi_{p}(\textbf{k}_{\bot})$ [$\chi_{s}(\textbf{k}_{\bot})$] will partially cancel out the gain effect of $\rho^{(0)}_{23}$ [$\rho^{(0)}_{24}$] as shown in Eqs.~(\ref{eq23}a) and (\ref{eq23}b). However, Fig.~\ref{fig:lin} shows that together with the nonlinear gain from the FWM process, the loss can be eliminated or even over-compensated.

 \begin{figure}[t]
\centering
\includegraphics[width=0.8\columnwidth]{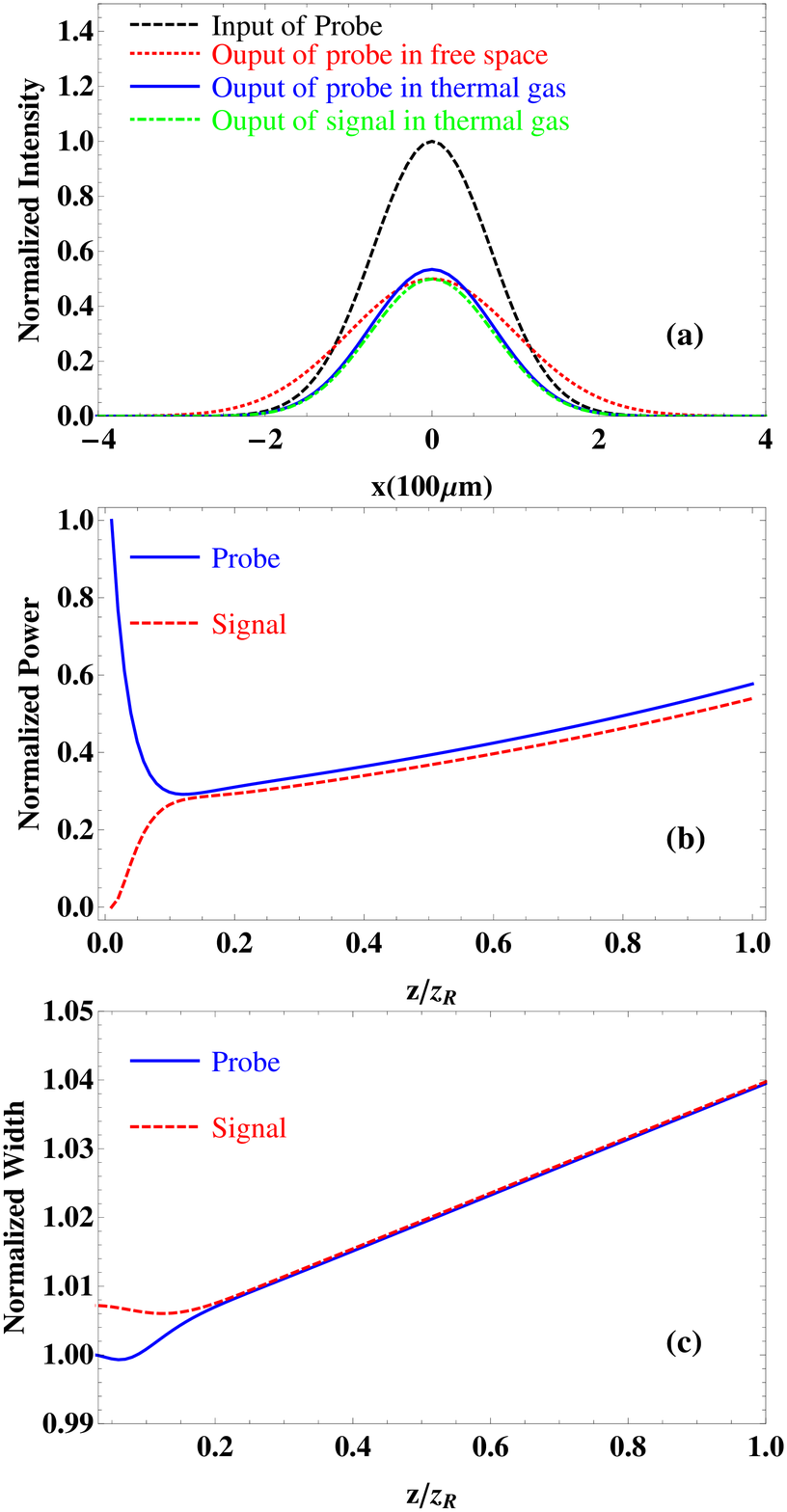}
\caption{(Color online)  (a) Transverse intensity profile of a two-dimensional Gaussian probe beam after propagation over one Rayleigh length. The figure shows the one-dimensional slice $y=0$. The different curves compare the profile of the probe beam in free space with those of the probe and FWM signal beams in the thermal medium. The intensities of all beams are scaled such that the peak intensity of the input probe field also shown in the figure is normalized. It can be seen that the power of the input beam is distributed between the probe and FWM signal beam, and that in the thermal medium, diffraction is significantly reduced compared to the free space case.
(b) and (c) show the normalized power and width of the probe and FWM signal as a function of the propagation distance. Note that the starting point for $z$ in (c) is $z=0.03z_{R}$, as at small propagation distance, the signal field is very weak as shown in (b), such that the width cannot easily be extracted. Parameters are the same as in Fig.~\ref{fig:lin}. } 
\label{fig:prop-1d}
\end{figure}

\subsection{\label{sec:dynamics}Propagation dynamics}

In Sec.~\ref{sec:sus}, we have discussed the essential mechanism for the elimination of the paraxial diffraction of arbitrary spatial profiles of both probe and signal fields, after the FWM process has evolved into the balanced state. Eqs.~(\ref{eq3}) and (\ref{eq23}) show that each wave vector component of the probe field is proportionally transfered onto the signal field and vice versa. Therefore, eventually the incident image carried by the probe field is copied to the signal via the interplay of the FWM processes. In this process, the relative intensities of the output probe and signal fields depends on their coupling strengths to the atomic medium.

To study the propagation dynamics, we numerically solve Eqs.~(\ref{eq3}) and (\ref{susceptibilities}). In the first step,  we start with a Gaussian probe field
\begin{align}
\Omega_{p}(\text{r}_{\bot},z=0)=\Omega_{p0}e^{-\frac{(x^2+y^2)}{2\text{w}^2_{p0}}} 
\label{eq25} 
\end{align}
The initial width of the probe beam is $\text{w}_{p0}=100\mu$m. Results are shown in Fig.~\ref{fig:prop-1d} for the $y=0$ subspace. In (a), after a propagation over one Rayleigh length $z_{R}=2\pi\text{w}^{2}_{p0}/\lambda_{p}=7.905$cm, in free space the transmitted probe beam is broadened to $\text{w}_{p}(z=z_{R})=\sqrt{2}\text{w}_{p0}$ due to paraxial diffraction. Accordingly, the intensity is decreased by factor of $1/2$. In the thermal vapor using the setup proposed here,  the width of the outgoing probe field remains almost constant $\text{w}_{p}(z=z_{R})\approx1.0395\text{w}_{p0}$, while the intensity is reduced to $\Omega_{p}({z=z_{R}})\approx0.537\Omega_{p0}$. 
The output of the generated FWM signal field has a Gaussian spatial structure with width $\text{w}_{s}(z=z_{R})\approx 1.0397\text{w}_{p0}$ and relative intensity $\Omega_{s}({z=z_{R}})\approx 0.502\Omega_{p0}$ similar to the output probe field. 
In Fig.~\ref{fig:prop-1d}(b) and (c), the relative widths and powers of both  fields are shown as a function of the propagation distance $z$. In the first stage of the propagation, the forward FWM process is dominant and transfers energy from the probe to the signal field. At about $z_{B}\approx 0.15z_{R}$, the forward and backward FWM processes are balanced. At this point, the normalized width of both signal and probe field has changed by less than one percent from the width of the initial probe field. In this initial propagation part, linear single-photon absorption dominates, and the total power contained in both signal and probe fields together is attenuated. Beyond $z>z_{B}$, the nonlinear gain over-compensates the linear loss, such that the probe and signal intensity grows slightly. This growth in power is accompanied by a small growth in the widths of signal and probe fields, which is related to  higher-order diffraction $\sim O(\text{k}^4_{\bot})$ in the linear and nonlinear dispersions. These are 
relevant, as the frequency spectrum of the incident pulse contains contributions at wave vectors outside the central region $\textbf{k}_{\bot}\ll\text{k}_{1}$.

\begin{figure}[t]
\centering
\includegraphics[width=0.8\columnwidth]{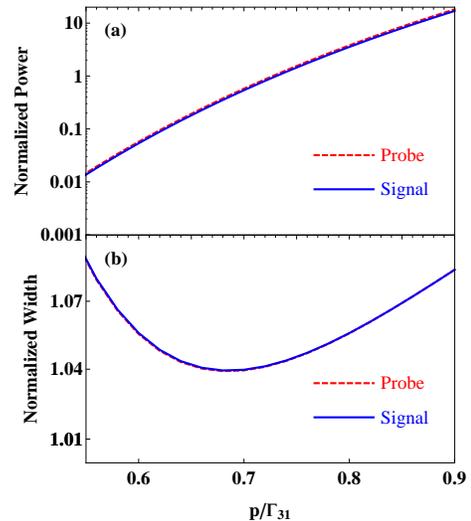}
\caption{(Color online)  The normalized output powers (a) and widths (b) of a Gaussian probe and the generated FWM signal field. Results are shown after propagation over one Rayleigh length in the thermal medium, as a function of the amplitude of the two-way incoherent pump $p$. If can be seen that already after a short propagation distance, the output probe and FWM signal fields widths and powers approach each other. Afterwards, the powers increase slowly over the residual propagation distance due to overall gain. Throughout the whole propagation, the widths of the two beams remains essentially unchanged, as can be seen from the small scale on the y-axis of (b).  Parameters are the same as in Fig.~\ref{fig:lin}. } 
\label{fig:power-1d}
\end{figure}

\begin{figure}[t]
\centering
\includegraphics[width=9cm]{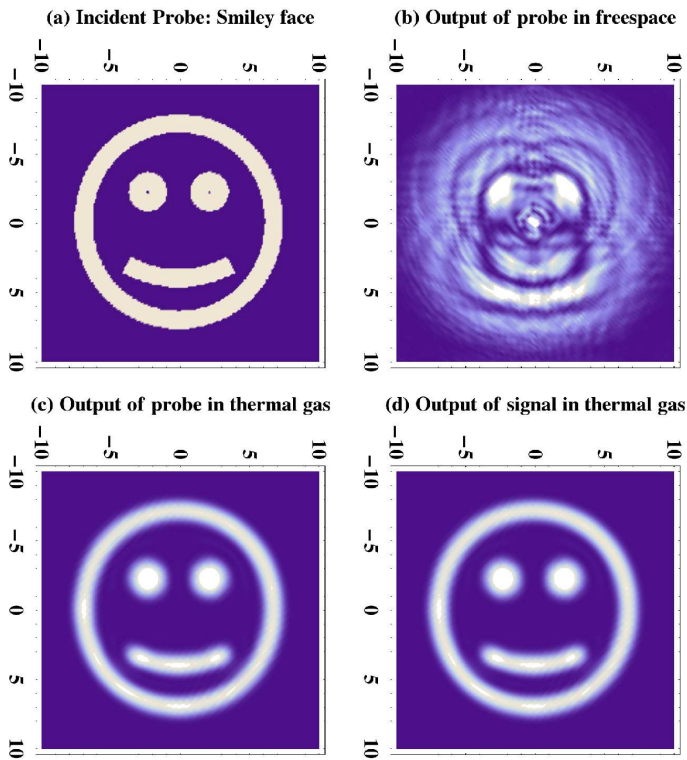}
\caption{(Color online)  Diffraction-less propagation and frequency conversion of arbitrary images. The original two-dimensional image encoded onto the transverse profile of the probe beam is shown in (a). In free space, the image is severely distorted after propagation over one Rayleigh length, as shown in  (b). In a suitably tailored thermal medium, the probe beam profile is preserved, see c). Additionally, the same profile is converted onto the signal field via the FWM process. All axis labels are given in units of $100\mu m$. Other parameters are as in Fig.~\ref{fig:lin}. } 
\label{fig:prop-2d}
\end{figure}

\subsection{\label{sec:pumpdynamics}Effect of the incoherent pump on the propagation dynamics}
To further analyze the effect of the incoherent pump beam on the propagation dynamics, we calculated the output power and width of both probe and signal fields as a function of pump intensity. Results are shown in Fig.~\ref{fig:power-1d} for a Gaussian probe field. We find that while the output power of the two fields can be controlled via the pump field over a large range, the output width only weakly depends on the pump power. This suggests that possible inhomogeneities in the incoherent pump field, which could, for example, arise due to absorption in the medium, will not significantly affect the outgoing spatial beam profiles and the diffraction-less propagation in the medium. We further find that the balanced distribution of the outgoing power between signal and probe field is preserved over a broad range of pump strengths as well. 
 
While the initial purpose of the incoherent pump field was to induce atomic coherences via the control fields in zeroth order of the probe and signal field, we find that it has further less obvious positive effects on the system dynamics.
First, the single-photon absorption without pump can be  compensated or even turned into gain by the combination of reduced linear absorption and nonlinear gain from the atomic coherences. This suggests that the FWM process, which sensitively depends on the relative intensities of all applied fields,  equilibrates in a rather short propagation distance. Consequently, not only the efficiency of the FWM process is significantly improved, but also the broadening of both probe and signal fields due to residual diffraction before the FWM equilibrium is reached is considerably decreased. 
Second, the incoherent pump further alleviates the demand for strong collision rates to achieve the Dicke limit Eq.~(\ref{dickelimit}), as $\gamma_{c1}$ is enhanced by the pump rate $p$. 
Third, since $\text{k}_{1}$ defined in Eq.~(\ref{k1}) which sets the transverse wave number scales in Eq.~(\ref{expand}) and Figs.~\ref{fig:lin} and~\ref{fig:nonlin} grows rapidly as the incoherent pump increases, probe and signal  fields with a larger transverse bandwidth, and in turn smaller spatial size can be propagated through the medium without diffraction in the thermal medium.

\subsection{\label{sec:2d}Diffraction-less propagation of arbitrary images}

In order to demonstrate that our setup can operate on arbitrary images encoded onto the probe field,  we finally propagated a two-dimensional image shown in Fig.~\ref{fig:prop-2d} through the thermal atom vapor. The incident image is represented by an array of values ``0" and ``1" in the transverse plane, which leads to the somewhat rough edges visible in Fig.~\ref{fig:prop-2d}(a). In free space, as expected we find that the image is totally distorted after propagating one Rayleigh length, see (b). But in the thermal vapor, the image encoded onto the  probe beam is well preserved due to the cancellation of the diffraction due to atomic motion. In comparison to the initial image, the sharp edges have been smoothed out, as they contain high-frequency transverse wave vector components  outside the central susceptibility area $\textbf{k}_{\bot}\ll\textbf{k}_{1}$. At the same time, the image is transfered to the signal field via the FWM process, see (d).

\section{\label{sec:summary}Discussions and Conclusion}

We have proposed a scheme to realize diffraction-less image propagation and frequency conversion based on four-wave mixing in a thermal atomic vapor.  Phase matching is achieved despite the thermal motion, which is exploited to achieve diffraction-less image propagation. In order to compensate the usually inevitable absorption, we applied a two-way incoherent field. It on the one hand modifies the linear and non-linear absorption such that overall, the probe and signal fields propagate without attenuation. But additionally, the pump field leads to a rapid equilibration of the FWM processes, which sensitively depends on the intensities of all applied fields. Thereby, an image frequency conversion with negligible diffraction in the initial transient dynamics is achieved. Furthermore, the additional pump field allows to enter the desired Dicke regime already at lower densities, which is favorable for experimental implementations. Finally, the pump field increases the transverse wave number bandwidth which can be 
propagated without diffraction through the thermal medium. We have shown that our method is capable of transmitting and frequency-converting complex two-dimensional images without diffraction, requiring only short propagation distances of less than one Rayleigh length.

Throughout the manuscript, we used parameters of the hyperfine structure of $^{87}\text{Rb}$ for the results, which is an atomic species routinely used in laboratories and therefore well suited for proof-of-principle experiments. But our setup can also be generalized to different systems, e.g., to achieve conversion to other frequencies. The main considerations which led us to $^{87}\text{Rb}$ are as follows. A major restriction on the parameters arises from the fact that the Doppler effect effectively reduces the coupling between the thermal vapor and the laser fields. Therefore, usually a relatively high atomic density ($>10^{12}~\text{cm}^{-3}$) is required to achieve sufficient medium response to cancel the diffraction, which can lead to unwanted effects such as additional non-linear processes or collective interactions~\cite{fleischhaker}. In order to reduce the required atomic density, the population in the upper state $|5\rangle$ should be minimized, which  acts only as an auxiliary state in order to 
effectively convert the incoherent two-way pump field into a one-way pumping. For this, the spontaneous decay rate $\Gamma_{52}$ should ideally be much larger than $\Gamma_{51}$. Additionally, the couplings between atoms and the probe and signal fields should be as strong as possible to further reduce the atomic density, such that large dipole moment for $\vec\mu_{31}$ and $\vec\mu_{41}$ are desirable. Lastly, the simultaneous elimination of diffraction for both fields in the same thermal vapor demands for matched coupling strengths, which further limits the choice of suitable  $\vec\mu_{31}$ and $\vec\mu_{41}$. 

We recently became aware of related work by I. Katzir et al., who independently of us developed another method to eliminate paraxial diffraction using a FWM process~\cite{katzir}. Interestingly, while the level schemes at first sight look rather similar, the involved physical processes are entirely different. The setup by Katzir et al. and ours share the main principle of operating in the transverse wave vector space $\textbf{k}_{\bot}$, such that diffraction of arbitrary images encoded onto the probe beam can be manipulated. But in contrast to the present work, Katzir et al. do not rely on the dependence of $\textbf{k}_{\bot}$ induced by the thermal atom motion, but instead exploit a new mechanism based on the phase-matching condition of a suitably tailored 4WM process. In their scheme, a probe field with arbitrary transverse profile and a plane-wave control field interact with a three-level $\Lambda$ system in EIT configuration. The control field further acts as a pump field to form an active Raman gain (ARG)
~\cite{wang,deng3} process, which leads to generation of a {\it conjugate} FWM signal field. Since for each wave component the produced conjugate signal field has an opposite phase dependence on the transverse wavevector as compared to the probe field, the diffraction for both fields can be tuned such that they cancel each other. As a result, the spatial profile of the probe field is well preserved, and copied to the signal field as in our proposal, while avoiding absorption via the non-linear gain. 
 
The two related yet different setups clearly highlight how versatile and intriguing the light propagation through atomic vapors becomes in particular if setups beyond the standard 3-level EIT setup are considered, and pave the way for further studies operating in the transverse wave vector space applicable to arbitrary probe images.

We thank O. Firstenberg for fruitful discussions. Funding by the German Science Foundation (DFG, Sachbeihilfe EV 157/2-1) is gratefully acknowledged.

\appendix

\numberwithin{equation}{section}

\section{\label{sec:appendix}Derivation of the susceptibilities}
In this Appendix, details of the derivation of the linear and nonlinear susceptibilities of the thermal atomic medium are summarized. 
 In order to describe the interaction between the laser fields and the thermal atomic medium, we follow~\cite{firstenberg2} and introduce a generalized density matrix distribution function including position ($\textbf{r}$) and velocity $(\textbf{v})$ information, 
\begin{equation} 
\rho(\textbf{r},\textbf{v},t)=\sum\limits_{i} \rho^{i}(t)\delta(\textbf{r}-\textbf{r}_{i}(t))\delta(\textbf{v}-\textbf{v}_{i}(t)),
\label{eq1}
\end{equation}
where $\rho^{i}(t)$ is the internal electronic density matrix for the $i$-th atom. $\rho(\textbf{r},\textbf{v},t)$ can then be interpreted as the probability density to find an atom characterized by internal density matrix $\rho(t)$ at position $\textbf{r}$ and with velocity $\textbf{v}$. The equation of motion of the system can be derived as
\begin{align}
\frac{\partial \rho(\textbf{r},\textbf{v},t)}{\partial t} &=- \frac{i}{\hbar} [H, \rho(\textbf{r},\textbf{v},t)] -L\rho(\textbf{r},\textbf{v},t) - \textbf{v}\cdot\frac{\partial \rho(\textbf{r},\textbf{v},t)}{\partial \textbf{r}}  \nonumber \\ 
&\quad -\gamma_{c}\{\rho(\textbf{r},\textbf{v},t) - R(\textbf{r},t)F(\textbf{v})\},
\label{eq2}
\end{align}
where $H$ is the Hamiltonian of the system, and $L\rho$ represents the incoherent internal dynamics including spontaneous decay, dephasing and the incoherent pumping field. Next to these quantum mechanical contributions,  the external classic motion leads to additional terms with $\gamma_{c}$ as the collision rate between atoms, 
\begin{align}
R(\textbf{r},t) = \int\rho(\textbf{r},\textbf{v},t)d\textbf{v}\,,
\end{align}
the number of atoms with density matrix $\rho(t)$ per unit volume at $\textbf{r}$, and 
\begin{align}
F(\textbf{v})=\frac{1}{(\sqrt{\pi}\text{v}_{\text {th}})^3 }\:e^{-\text{v}^2/\text{v}_{\text{th}}^2}
\end{align}
is the Boltzmann distribution with $\text{v}_{\text{th}}=\sqrt{2k_{b}T/m}$ being the thermal velocity.  Specifically, the third term addresses the thermal motion, and the last term is related to the phase-changing collisions between atoms.

To determine the medium susceptibilities, from Eq.~(\ref{eq2}), we can derive the corresponding density matrix equations of motion for the FWM process in the undepleted-pump limit to give
\begin{subequations}
\label{eq4}
\begin{align}
&\left( \frac{\partial}{\partial t} + \textbf{v}\cdot\frac{\partial}{\partial \textbf{r}}-i\Delta_{s}+i\textbf{k}_{s}\cdot\textbf{v}+\frac{p(\textbf{r})}{2}+\frac{\Gamma_{4}}{2}+\gamma_{c}\right)\rho_{41} \nonumber \\
&= i\Omega_{s}(\textbf{r},t)(\rho_{11}-\rho_{44})-i\Omega_{p}(\textbf{r},t)\rho_{43}+i\Omega_{c2}(\textbf{r},t)\rho_{21} \nonumber \\
&\quad+\gamma_{c}R_{41}(\textbf{r},t)F(\textbf{v})\,,       \\
&\left(\frac{\partial}{\partial t} + \textbf{v}\cdot\frac{\partial}{\partial \textbf{r}}-i\Delta_{p}+i\textbf{k}_{p}\cdot\textbf{v}+\frac{p(\textbf{r})}{2}+\frac{\Gamma_{3}}{2}+\gamma_{c}\right)\rho_{31} \nonumber \\
&= i\Omega_{p}(\textbf{r},t)(\rho_{11}-\rho_{33})-i\Omega_{s}(\textbf{r},t)\rho_{34} +i\Omega_{c1}(\textbf{r},t)\rho_{21}   \nonumber \\
&\quad+\gamma_{c}R_{31}(\textbf{r},t)F(\textbf{v})\,,       \\
&\left( \frac{\partial}{\partial t} + \textbf{v}\cdot\frac{\partial}{\partial \textbf{r}}-i\Delta+i\Delta\textbf{k}\cdot\textbf{v}+\frac{p(\textbf{r})}{2}+\gamma_{21}+\gamma_{c}\right)\rho_{21} \nonumber \\
&= i\Omega_{c1}^{*}(\textbf{r},t)\rho_{31}+i\Omega_{c2}^{*}(\textbf{r},t)\rho_{41}-i\Omega_{p}(\textbf{r},t)\rho_{23} \nonumber \\
&\quad-i\Omega_{s}(\textbf{r},t)\rho_{24} +\gamma_{c}R_{21}(\textbf{r},t)F(\textbf{v})\,.
\end{align}
\end{subequations}
Note we have simplified the notation of $\rho_{ij}(\textbf{r},\textbf{v},t)$ to $\rho_{ij}$. Further, 
\begin{align}
\Delta &= \Delta_{p}-\Delta_{c1}\quad \textrm{and}\\
\Delta\textbf{k} &=\textbf{k}_{p}-\textbf{k}_{c1}
\end{align}
are the two-photon detuning and the wavevector difference between $\Omega_p$ and $\Omega_{c1}$. The phase-matching conditions are
\begin{align}
\Delta_{s}&=\Delta_{p}-\Delta_{c1}+\Delta_{c2}\,,\\
\textbf{k}_{s}&=\textbf{k}_{p}-\textbf{k}_{c1}+\textbf{k}_{c2}\,.
\end{align}
Here, $\Delta_{i}$ and $\textbf{k}_{i}$ are the detuning and wave vector of the field with Rabi frequency $\Omega_{i} (\textbf{r},t)$ ($i\in\{p, s, c1, c2\}$), $p(\textbf{r})$ is the incoherent pump rate, $\Gamma_{i}=\sum_{k}\Gamma_{ik}$ the total spontaneous emission rate from state $|i\rangle$ with $\Gamma_{ik}$ the partial one from $|i\rangle$ to $|k\rangle$, and $\gamma_{21}$ denotes the ground state dephasing.

For simplicity, we focus on the case in which the two control fields and the incoherent pump can be treated as plane waves, such that $\Omega_{c1}(\textbf{r})=\Omega_{c1}$, $\Omega_{c2}(\textbf{r})=\Omega_{c2}$, and $p(\textbf{r})=p$. Moreover, we expand the system to leading order of the weak probe and FWM signal fields. In zeroth order, the steady-state density-matrix distribution function is obtained as 
\begin{equation}
\rho^{(0)}_{ij}(\textbf{r},\textbf{v},t\rightarrow\infty)=n_{0}\rho^{(0)}_{ij}F(\textbf{v})
\label{eq5}
\end{equation}
where $n_{0}$ is the atomic density, and $\rho^{(0)}_{ij}$ is the zero-order density-matrix element of an atom at rest. Note that $\rho^{(0)}_{ij}$ can easily be calculated from the steady-state master equations by setting $\Omega_{p(s)}=0$.

The first-order equations of motion then follow from Eq.~(\ref{eq4}) as
\begin{subequations}
\label{eq6}
\begin{align}
&\left( \frac{\partial}{\partial t} + \textbf{v}\cdot\frac{\partial}{\partial \textbf{r}}-i\Delta_{s}+i\textbf{k}_{s}\cdot\textbf{v}+\frac{p}{2}+\frac{\Gamma_{4}}{2}+\gamma_{c}\right) \rho^{(1)}_{41} \nonumber \\
&= i\Omega_{s}(\textbf{r},t)(\rho^{(0)}_{11}-\rho^{(0)}_{44})n_{0}F(\textbf{v})-i\Omega_{p}(\textbf{r},t)\rho^{(0)}_{43}n_{0}F(\textbf{v}) \nonumber \\
&\quad+i\Omega_{c2}\rho^{(1)}_{21}+\gamma_{c}R^{(1)}_{41}(\textbf{r},t)F(\textbf{v})\,,       \\[2ex]
&\left(\frac{\partial}{\partial t} + \textbf{v}\cdot\frac{\partial}{\partial \textbf{r}}-i\Delta_{p}+i\textbf{k}_{p}\cdot\textbf{v}+\frac{p}{2}+\frac{\Gamma_{3}}{2}+\gamma_{c}\right)\rho^{(1)}_{31} \nonumber \\
&= i\Omega_{p}(\textbf{r},t)(\rho^{(0)}_{11}-\rho^{(0)}_{33})n_{0}F(\textbf{v})-i\Omega_{s}(\textbf{r},t)\rho^{(0)}_{34}n_{0}F(\textbf{v}) \nonumber \\
&\quad+i\Omega_{c1}\rho^{(1)}_{21} +\gamma_{c}R^{(1)}_{31}(\textbf{r},t)F(\textbf{v})\,,       \\[2ex]
&\left(\frac{\partial}{\partial t} + \textbf{v}\cdot\frac{\partial}{\partial \textbf{r}}-i\Delta+i\Delta\textbf{k}\cdot\textbf{v}+\frac{p}{2}+\gamma_{21}+\gamma_{c}\right) \rho^{(1)}_{21} \nonumber \\
&= i\Omega_{c1}^{*}\rho^{(1)}_{31}+i\Omega_{c2}^{*}\rho^{(1)}_{41}-i\Omega_{p}(\textbf{r},t)\rho^{(0)}_{23}n_{0}F(\textbf{v}) \nonumber \\
&\quad-i\Omega_{s}(\textbf{r},t)\rho^{(0)}_{24}n_{0}F(\textbf{v}) +\gamma_{c}R^{(1)}_{21}(\textbf{r},t)F(\textbf{v})\,.
\end{align}
\end{subequations}
From Eqs.~(\ref{eq6}), the general solution for $R^{(1)}_{ij}(\textbf{k},\omega)$ can be obtained by Fourier transforming from $(\textbf{r},t)$  coordinates to $(\textbf{k},\omega)$ space, and solving the resulting algebraic equations for $\rho^{(1)}_{ij}(\textbf{k},\omega)$. The result for $\rho^{(1)}_{ij}(\textbf{k},\omega)$ is integrated over velocity to give a set of equations for $R^{(1)}_{ij}(\textbf{k},\omega)$. Solving the resulting equations leads to an  expression for $R^{(1)}_{ij}(\textbf{k},\omega)$. While $R^{(1)}_{ij}(\textbf{k},\omega)$ can be obtained from these straightforward procedures, its final expressions are too complicated to gain deeper physical understanding of the FWM process. 

Therefore, we apply a different procedure. To this end, we first integrate Eq.~(\ref{eq6}c) over velocity and obtain
\begin{align}
&\left(\frac{\partial}{\partial t} -i\Delta+\frac{p}{2}+\gamma_{21}\right)R^{(1)}_{21}(\textbf{r},t) +\left(\frac{\partial}{\partial \textbf{r}}+i\Delta\textbf{k}\right))\textbf{J}_{21}(\textbf{r},t)\nonumber \\
&= i\Omega_{c1}^{*}R^{(1)}_{31}(\textbf{r},t)+i\Omega_{c2}^{*}R^{(1)}_{41}(\textbf{r},t)-in_{0}\Omega_{p}(\textbf{r},t)\rho^{(0)}_{23} \nonumber \\
&\quad-in_{0}\Omega_{s}(\textbf{r},t)\rho^{(0)}_{24} \,,
\label{eq7}
\end{align}
where we have defined the current density of the density-matrix distribution function
\begin{equation}
\textbf{J}_{ij}(\textbf{r},t)=\int\textbf{v}\rho^{(1)}_{ij}(\textbf{r},\textbf{v},t)d\textbf{v}  \,.  \\
\label{eq8}
\end{equation}
For dominating
\begin{align}
\gamma=\gamma_{c}+\frac{p}{2}+\gamma_{21}-i\Delta \,,
\end{align}
we can expand $\rho(\textbf{r},\textbf{v},t)$ to first order in $\gamma$  as
\begin{align}
\label{eq9}
 \rho^{(1)}_{21}(\textbf{r},\textbf{v},t)&=
\rho^{(1,0)}_{21}(\textbf{r},\textbf{v},t)+\frac{1}{\gamma}\rho^{(1,1)}_{21}(\textbf{r},\textbf{v},t) \nonumber \\
&=R^{(1)}_{21}(\textbf{r},t)F(\textbf{v})+\frac{1}{\gamma}\rho^{(1,1)}_{21}(\textbf{r},\textbf{v},t)
\end{align}
to obtain
\begin{subequations}
\label{eq10}
\begin{align}
0&=\int\textbf{v}\rho^{(1,0)}_{21}(\textbf{r},\textbf{v},t)d\textbf{v}\,, \\
\textbf{J}_{21}(\textbf{r},t)&=\frac{1}{\gamma}\int\textbf{v}\rho^{(1,1)}_{21}(\textbf{r},\textbf{v},t)d\textbf{v} \,.
\end{align}
\end{subequations}
Next, we multiply Eq.~(\ref{eq6}c) by $\textbf{v}$, and integrate over velocity. Expanding to the leading term in $\gamma$ and inserting the relations in Eq.~(\ref{eq10}), we obtain the following equation for $\textbf{J}_{21}(\textbf{r},t)$:
\begin{align}
\textbf{J}_{21}(\textbf{r},t)&=-D\left( \frac{\partial}{\partial\textbf{r}}+i\Delta\textbf{k}\right) R^{(1)}_{21}(\textbf{r},t)\nonumber\\
&\quad+i\left( \frac{\Omega_{c1}^{*}}{\gamma}\textbf{J}_{31}(\textbf{r},t)+\frac{\Omega_{c2}^{*}}{\gamma}\textbf{J}_{41}(\textbf{r},t)\right)\,.
\label{eq11} 
\end{align}
Here, $D=v_{\text{th}}^{2}/\gamma$, and to derive Eq.~(\ref{eq11}) we have used the relation~\cite{firstenberg2}
\begin{align}
\int\textbf{v}^2\frac{\partial}{\partial\textbf{r}}R^{(1)}_{21}(\textbf{r},t)F(\textbf{v})d\textbf{v}=\text{v}_{\text{th}}^2\frac{\partial}{\partial\textbf{r}}R^{(1)}_{21}(\textbf{r},t)\,.
\label{eq12} 
\end{align}
Inserting Eq.~(\ref{eq11}) into Eq.~(\ref{eq7}), we find
\begin{align}
&\left( \frac{\partial}{\partial t} -i\Delta+\frac{p}{2}+\gamma_{21}-D\big(\frac{\partial}{\partial\textbf{r}}+i\Delta\textbf{k}\big)^2\right) R^{(1)}_{21}(\textbf{r},t)\nonumber \\
&= i\Omega_{c1}^{*}R^{(1)}_{31}(\textbf{r},t)+i\Omega_{c2}^{*}R^{(1)}_{41}(\textbf{r},t)\nonumber \\
&\quad-in_{0}\Omega_{p}(\textbf{r},t)\rho^{(0)}_{23}-in_{0}\Omega_{s}(\textbf{r},t)\rho^{(0)}_{24}\nonumber \\
&\quad-i\left(\frac{\partial}{\partial\textbf{r}}+i\Delta\textbf{k}\right)\cdot\left(\frac{\Omega_{c1}^{*}}{\gamma}\textbf{J}_{31}(\textbf{r},t)+\frac{\Omega_{c2}^{*}}{\gamma}\textbf{J}_{41}(\textbf{r},t)\right)\,.
\label{eq13} 
\end{align}
In Eq.~(\ref{eq13}), the last terms containing $\textbf{J}_{31}(\textbf{r},t),\textbf{J}_{41}(\textbf{r},t)$ can be neglected if $|\Omega_{c1}|,|\Omega_{c2}|\ll|\gamma|$, which is typically satisfied in relevant setups. Furthermore, even when this condition is not valid, these terms can still be neglected if both the spatial variations $\partial/\partial\textbf{r}$ and $\Delta\textbf{k}$ remain in the transverse plane, perpendicular to $\textbf{k}_{p}$ and $\textbf{k}_{s}$~\cite{firstenberg2}. Then Eq.~(\ref{eq13}) reduces to
\begin{align}
&\left( \frac{\partial}{\partial t} -i\Delta+\frac{p}{2}+\gamma_{21}-D\big(\frac{\partial}{\partial\textbf{r}}+i\Delta\textbf{k}\big)^2\right)R^{(1)}_{21}(\textbf{r},t)\nonumber \\
&= i\Omega_{c1}^{*}R^{(1)}_{31}(\textbf{r},t)+i\Omega_{c2}^{*}R^{(1)}_{41}(\textbf{r},t)\nonumber \\
&\quad-in_{0}\Omega_{p}(\textbf{r},t)\rho^{(0)}_{23}-in_{0}\Omega_{s}(\textbf{r},t)\rho^{(0)}_{24}\,.
\label{eq14} 
\end{align}
In the slowly-varying envelope approximation (SVEA), the temporal and spatial variations of the envelope of probe and signal fields are assumed to be much smaller than the decoherence rate and the wave number. For the spatio-temporal evolution of the density-matrix distribution function $R^{(1)}_{31}(\textbf{r},t),R^{(1)}_{41}(\textbf{r},t)$, SVEA leads to the conditions
\begin{subequations}
\label{eq15}  
\begin{align}
&\bigg|\frac{\partial}{\partial t}+\textbf{v}\cdot\frac{\partial}{\partial\textbf{r}}\bigg|\ll\bigg|\frac{p}{2}+\frac{\Gamma_{4}}{2}-i\Delta_{s}+i\textbf{k}_{s}\cdot\textbf{v}\bigg|\,, \\
&\bigg|\frac{\partial}{\partial t}+\textbf{v}\cdot\frac{\partial}{\partial\textbf{r}}\bigg|\ll\bigg|\frac{p}{2}+\frac{\Gamma_{3}}{2}-i\Delta_{p}+i\textbf{k}_{p}\cdot\textbf{v}\bigg| \,.
\end{align} 
\end{subequations}
Neglecting the temporal and spatial variations and only taking the dominant part of $\rho^{(1)}_{21}(\textbf{r},\textbf{v},t)=R^{(1)}_{21}(\textbf{r},t)F(\textbf{v})$ in Eqs.~(\ref{eq6}a) and (\ref{eq6}b), by integrating over velocity, we obtain expressions for $R^{(1)}_{31}(\textbf{r}$,$\textbf{v},t)$ and $R^{(1)}_{41}(\textbf{r},\textbf{v},t)$, 
\begin{subequations}
\label{eq16}
\begin{align}
R^{(1)}_{41}(\textbf{r},t)&=iK_{41}\left( n_{0}(\rho^{(0)}_{11}-\rho^{(0)}_{44})\Omega_{s}(\textbf{r},t)
-n_{0}\rho^{(0)}_{43}\Omega_{p}(\textbf{r},t) \right. \nonumber\\
&\left. \quad+\Omega_{c2}R^{(1)}_{21}(\textbf{r},t)\right) \,,\\
R^{(1)}_{31}(\textbf{r},t)&=iK_{31}\left( n_{0}(\rho^{(0)}_{11}-\rho^{(0)}_{33})\Omega_{p}(\textbf{r},t)
-n_{0}\rho^{(0)}_{34}\Omega_{s}(\textbf{r},t) \right. \nonumber\\
&\left. \quad+\Omega_{c1}R^{(1)}_{21}(\textbf{r},t)\right)\,,
\end{align}
\end{subequations} 
where $K_{41}$ and $K_{31}$ are defined as 
\begin{subequations}
\label{eq17}
\begin{align}
K_{31}&=\frac{iG_{31}}{1-i\gamma_{c}G_{31}} \,, \\
K_{41}&=\frac{iG_{41}}{1-i\gamma_{c}G_{41}}\,, \\
G_{31}&=\int\frac{F(\textbf{v})}{\Delta_{s}-\textbf{k}_{s}\cdot\textbf{v}+i(\frac{p}{2}+\frac{\Gamma_{3}}{2}+\gamma_{c})}d^{3}\textbf{v} \,, \\
G_{41}&=\int\frac{F(\textbf{v})}{\Delta_{p}-\textbf{k}_{p}\cdot\textbf{v}+i(\frac{p}{2}+\frac{\Gamma_{4}}{2}+\gamma_{c})}d^{3}\textbf{v}\,.
\end{align}
\end{subequations}
Near the one-photon resonance with small detunings $\Delta_{p}\ll p/2+\Gamma_{3}+\gamma_{c}$ and $\Delta_{s}\ll p/2+\Gamma_{4}+\gamma_{c}$,  the imaginary parts of $K_{31}$ and $K_{41}$ are much smaller than their respective real parts, and we neglect the imaginary parts in the following.

Following the paraxial approximation, we separate the transverse and longitudinal ($z$) coordinates, $\textbf{r}\rightarrow(\textbf{r}_{\bot},z)$, and neglect changes along the propagation direction, $\partial/\partial\textbf{r}\rightarrow\partial/\partial\textbf{r}_{\bot}$. Next, we Fourier transform Eqs.~(\ref{eq14}) and (\ref{eq16}) from $(\textbf{r}_{\bot},t)$ to $(\textbf{k}_{\bot},\omega)$, and obtain the final expressions for $R^{(1)}_{31}(\textbf{k}_{\bot},\omega)$ and $R^{(1)}_{41}(\textbf{k}_{\bot},\omega)$ by solving the  equations
\begin{widetext}
\begin{subequations}
\begin{align}
R^{(1)}_{31}(\textbf{k}_{\bot},z,\omega)&=iK_{31}n_{0}\Omega_{p}(\textbf{k}_{\bot},z,\omega)
\big[\rho^{(0)}_{11}-\rho^{(0)}_{33}+\frac{K_{31}\Omega^2_{c1}(\rho^{(0)}_{11}-\rho^{(0)}_{33})+i\Omega_{c1}\rho^{(0)}_{23}-K_{41}\Omega_{c1}\Omega^{*}_{c2}\rho^{(0)}_{43}}{i(\omega+\Delta)-\frac{p}{2}-\gamma_{21}-D(\textbf{k}_{\bot}+\Delta\textbf{k})^2-K_{31}\Omega^{2}_{c1}-K_{41}\Omega^{2}_{c2}}\big]\nonumber \\
&\quad+iK_{31}n_{0}\Omega_{s}(\textbf{k}_{\bot},z,\omega)
\big[-\rho^{(0)}_{34}+\frac{K_{41}\Omega_{c1}\Omega^{*}_{c2}(\rho^{(0)}_{11}-\rho^{(0)}_{44})+i\Omega_{c1}\rho^{(0)}_{24}-K_{31}\Omega^{2}_{c1}\rho^{(0)}_{34}}{i(\omega+\Delta)-\frac{p}{2}-\gamma_{21}-D(\textbf{k}_{\bot}+\Delta\textbf{k})^2-K_{31}\Omega^{2}_{c1}-K_{41}\Omega^{2}_{c2}}\big]\,, \\[2mm]
R^{(1)}_{41}(\textbf{k}_{\bot},z,\omega)&=iK_{41}n_{0}\Omega_{s}(\textbf{k}_{\bot},z,\omega)
\big[\rho^{(0)}_{11}-\rho^{(0)}_{44}+\frac{K_{41}\Omega^2_{c2}(\rho^{(0)}_{11}-\rho^{(0)}_{44})+i\Omega_{c2}\rho^{(0)}_{24}-K_{31}\Omega^{*}_{c1}\Omega_{c2}\rho^{(0)}_{34}}{i(\omega+\Delta)-\frac{p}{2}-\gamma_{21}-D(\textbf{k}_{\bot}+\Delta\textbf{k})^2-K_{31}\Omega^{2}_{c1}-K_{41}\Omega^{2}_{c2}}\big]\nonumber \\
&\quad+iK_{41}n_{0}\Omega_{p}(\textbf{k}_{\bot},z,\omega)
\big[-\rho^{(0)}_{43}+\frac{K_{31}\Omega^{*}_{c1}\Omega_{c2}(\rho^{(0)}_{11}-\rho^{(0)}_{33})+i\Omega_{c2}\rho^{(0)}_{23}-K_{41}\Omega^{2}_{c2}\rho^{(0)}_{43}}{i(\omega+\Delta)-\frac{p}{2}-\gamma_{21}-D(\textbf{k}_{\bot}+\Delta\textbf{k})^2-K_{31}\Omega^{2}_{c1}-K_{41}\Omega^{2}_{c2}}\big]\,. 
\end{align}
\label{eq18} 
\end{subequations}
\end{widetext}
For continuous wave fields, we can set $\omega=0$ in Eqs.~(\ref{eq18}). In the case of a small wavevector difference between $\Omega_{p}$ and $\Omega_{c1}$, $\Delta\textbf{k}$ can be neglected, i.e., $\Delta\textbf{k}=0$. 

Finally, we note that the propagation equations for the two probe and signal fields in momentum space can be written as 
\begin{subequations}
\label{prop-m}
\begin{align}
\left(\frac{\partial}{\partial z}+i\frac{\text{k}_{\bot}^{2}}{2\text{k}_{p}}\right)\Omega_{p}(\textbf{k}_{\bot},z)
&=i\frac{3\lambda_{p}^2\Gamma_{31}}{8\pi}R_{31}(\textbf{k}_{\bot},z)\,,\\
\left(\frac{\partial}{\partial z}+i\frac{\text{k}_{\bot}^{2}}{2\text{k}_{s}}\right)\Omega_{s}(\textbf{k}_{\bot},z)
&=i\frac{3\lambda_{s}^2\Gamma_{41}}{8\pi}R_{41}(\textbf{k}_{\bot},z)\,.
\end{align}
\end{subequations}
By comparing Eqs.~(\ref{prop-m}) with Eqs.~(\ref{eq3}), we can then find the expression for the linear susceptibilities $\chi_p,\chi_s$ and nonlinear susceptibilities $\chi_{ps}, \chi_{sp}$ given in Eqs.~(\ref{susceptibilities}).




\begin{thebibliography}{99}
\bibitem{shen} Y. R. Shen, The Principles of Nonlinear Optics (John Wiley and Sons, New York, 1984).
\bibitem{tewari1} S. P. Tewari and G. S. Agarwal, Phys. Rev. Lett. {\bf56}, 1811 (1986).
\bibitem{harris1} S. E. Harris, J. E. Field, and A. Imamo\v{g}lu, Phys. Rev. Lett. {\bf64}, 1107 (1990).
\bibitem{tewari2} S. P. Tewari and G. S. Agarwal, Phys. Rev. Lett. {\bf66}, 1797 (1991).
\bibitem{deng1} L. Deng, and M. G. Payne, Phys. Rev. Lett. {\bf91}, 243902 (2003). 
\bibitem{wu1} Y. Wu, J. Saldana, and Y. F. Zhu, Phys. Rev. A {\bf67}, 013811 (2003).
\bibitem{sorokin1} P. P. Sorokin, J. J. Wynne, and J. R. Lankard, App. Phys. Lett. {\bf22}, 342 (1973).
\bibitem{zibrov1} A. S. Zibrov, A. B. Matsko, and M. O. Scully, Phys. Rev. Lett. {\bf89}, 103601 (2002).


\bibitem{conversion}G. Radnaev,	Y. O. Dudin,	R. Zhao,	H. H. Jen,	S. D. Jenkins,	A. Kuzmich, T. A. B. Kennedy,     Nature Phys. {\bf 6},  894 (2010).
\bibitem{ding} D. -S. Ding, Z. -Y. Zhou, B. -S. Shi, X. -B. Zou, and G. -C. Guo, Phys. Rev. A {\bf85}, 053815 (2012).
\bibitem{liu} X. Liu, B. Kuyken, G. Roelkens, R Baets, R. M. Osgood Jr, and W. M. J. Green, Nature Photon. {\bf6}, 667 (2012).

\bibitem{4wm-dynamics}R. Fleischhaker and J. Evers, Phys Rev. A {\bf 78}, 051802(R) (2008).

\bibitem{durnin}J. Durnin, J. Opt. Soc. Am. A {\bf 4}, 651 (1987). 


\bibitem{berry1979}  M. V. Berry and N. L. Balazs, Am. J. Phy. {\bf47}, 264 (1979).
\bibitem{siviloglou2007} G. A. Siviloglou, J. Broky, A. Dogariu, and D. N. Christodoulides, Phys. Rev. Lett. {\bf99}, 213901 (2007).
\bibitem{durnin1987} J. Durnin, J. J. Miceli, Jr., and J. H. Eberly, Phys. Rev. Lett. {\bf58}, 1499 (1987).
\bibitem{nature2002} V. Garc\'{e}s-Ch\'{a}vez, D. McGloin, H. Melville, W. Sibbett, and K. Dholakia, Nature {\bf419}, 145 (2002).
\bibitem{mcgloin2005} D. McGloina and K. Dholakia, Contemporary Physics {\bf46}, 15 (2005). 
\bibitem{vega2000} J. C. Guti\'{e}rrez-Vega, M. D. Iturbe-Castillo, and S. Ch\'{a}vez-Cerda, Opt. Lett. {\bf25}, 1493 (2000).
\bibitem{zhang2012} P. Zhang, Y. Hu, T. Li, D. Cannan, X. Yin, R. Morandotti, Z. Chen, and X. Zhang, Phys. Rev. Lett {\bf109}, 193901 (2012).
\bibitem{bandres2004} M. A. Bandres, J. C. Guti\'{e}rrez-Vega, and S. Ch\'{a}vez-Cerda, Opt. Lett. {\bf29}, 44 (2004).

\bibitem{moseley}R. R. Moseley, S. Shepherd, D. J. Fulton, B. D. Sinclair, 
and M. H. Dunn, Phys. Rev. Lett. {\bf 74}, 670 (1995). 

\bibitem{bortman}D. Bortman-Arbiv, A. D. Wilson-Gordon, and H. Friedmann, Phys. Rev. A {\bf 58}, R3403 (1998). 

\bibitem{vengalatorre}M. Vengalattore and M. Prentiss, Phys. Rev. Lett. {\bf 95}, 
243601 (2005). 

\bibitem{tarhan}D. Tarhan, N. Postacioglu, and Ozgur E. Mustecaplioglu, 
Opt. Lett. {\bf 32}, 1038 (2007). 

\bibitem{truscott1999} A. G. Truscott, M. E. J. Friese, N. R. Heckenberg, and H. Rubinsztein-Dunlop, Phys. Rev. Lett. {\bf82}, 1438 (1999).
\bibitem{agarwal2000} R. Kapoor and G. S. Agarwal, Phys. Rev. A {\bf61}, 053818 (2000).
\bibitem{howell2009} P. K. Vudyasetu, D. J. Starling, and J. C. Howell, Phys. Rev. Lett. {\bf102}, 123602 (2009).
\bibitem{dey2009} T. N. Dey and G. S. Agarwal, Opt. Lett. {\bf34}, 3199 (2009).
\bibitem{dey2011} T. N. Dey and J. Evers, Phys. Rev. A {\bf84}, 043842 (2011).
\bibitem{zhang2013} L. Zhang, T. N. Dey, and J. Evers, Phys. Rev. A {\bf87}, 043842 (2013).
\bibitem{verma} O. N. Verma, L. Zhang, J. Evers and T. N. Dey, Phys. Rev. A {\bf88}, 013810 (2013).


\bibitem{firstenberg1} O. Firstenberg, M. Shuker, A. Ben-Kish, D. R. Fredkin, N. Davidson, and A. Ron, Phys. Rev. A {\bf76}, 013818 (2007).
\bibitem{firstenberg2} O. Firstenberg, M. Shuker, R. Pugatch, D. R. Fredkin, N. Davidson, and A. Ron, Phys. Rev. A {\bf77}, 043830 (2008).
\bibitem{firstenberg3} O. Firstenberg, M. Shuker, N. Davidson, and A. Ron, Phys. Rev. Lett. {\bf102}, 043601 (2009).
\bibitem{firstenberg4} O. Firstenberg, P. London, M. Shuker, A. Ron and N. Davidson, Nature Phys. {\bf5}, 665 (2009).
\bibitem{rmp}O. Firstenberg, M. Shuker, A. Ron, N. Davidson, Rev. Mod. Phys. {\bf 85}, 941 (2013).

\bibitem{zibrov2} A. S. Zibrov, M. D. Lukin, L. Hollberg, D. E. Nikonov, M. O. Scully, H. G. Robinson, and V. L. Velichansky, Phys. Rev. Lett {\bf76}, 3935(1996).
\bibitem{kapale1} K. T. Kapale, M. O. Scully, S. Y. Zhu, and M. S. Zubairy, Phys. Rev. A {\bf67}, 023804(2003).
\bibitem{wang1} L. G. Wang, S. Qamar, S.Y. Zhu, and M. S. Zubairy, Phys. Rev. A {\bf77}, 033833(2008).
\bibitem{obrien1} C. O'Brien and O. Kocharovskaya, Phys. Rev. Lett. {\bf107}, 137401(2011).
\bibitem{fleischhaker}R. Fleischhaker, T. N. Dey, and J. Evers, Phys. Rev. A {\bf 82}, 013815 (2010).
\bibitem{katzir} I. Katzir, O. Firstenberg and A. Ron, in preparation.
\bibitem{wang} L. J. Wang, A. Kuzmich and A. Dogariu, Nature {\bf406}, 227 (2000).
\bibitem{deng3} L. Deng and M. G. Payne, Phys. Rev. Lett. {\bf98}, 253902 (2007).

\end{thebibliography}
\end{document}